\documentclass[adp,fleqn]{w-art}
\usepackage{times}
\usepackage{w-thm}
\usepackage[]{graphicx}
\usepackage{amssymb}

\newcommand{\be}{\begin{equation}}
\newcommand{\ee}{\end{equation}}
\newcommand{\bea}{\begin{eqnarray}}
\newcommand{\eea}{\end{eqnarray}}

\begin{document}
\DOIsuffix{theDOIsuffix}
\Receiveddate{28 October 2005}
\Reviseddate{}
\Accepteddate{16 January 2006}
\Dateposted{}

\keywords{phantom cosmology, observational cosmology, singularities}
\subjclass[pacs]{98.80.Cq, 04.20.Jb, 98.80.Jk}



\title[Future Universe]{Future State of the Universe}

\author[M.P. D\c{a}browski]{Mariusz P. D\c{a}browski
  \footnote{E-mail:~\textsf{mpdabfz@sus.univ.szczecin.pl},
            Phone: +48\,91\,444\,1248,
            Fax: +48\,91\,444\,1226}}
\address{Institute of Physics, University of Szczecin,
Wielkopolska 15, 70-451 Szczecin, Poland}


\begin{abstract}
Following the observational evidence for cosmic acceleration which
may exclude a possibility for the universe to recollapse to a
second singularity, we review alternative scenarios of its future
evolution. Although the de Sitter asymptotic state is still an
option, some other asymptotic states which allow new types of
singularities such as Big-Rip (due to a phantom matter) and sudden
future singularities are also admissible and are reviewed in detail.
The reality of these singularities which comes from the
relation to observational characteristics of the universe
expansion are also revealed and widely discussed.
\end{abstract}


\maketitle

\tableofcontents


\section{Introduction}

It is generally agreed that we have now enough evidence for the
past Hot-Big-Bang universe. Its main observational support relies
on the following facts:

\begin{itemize}

\item The universe expands, i.e., all the galaxies move away from
each other according to the Hubble law \cite{hubble} and they experience
cosmological redshift $z$ according to the relation:
\bea
1+z &=& \frac{a(t_0)}{a(t_e)}~,
\eea
where $a(t_0)$ is the scale factor at the time of observation,
while $a(t_e)$ is the scale factor at the time of emission of
light by a galaxy.

\item Element abundance in the universe is: hydrogen $75 \%$,
helium $24 \%$, and other elements $1\%$. In particular, the amount
of helium is larger than it is possible to be produced in stars,
and the only solution to this problem is to assume that its abundance
is primordial \cite{gamow}.

\item Cosmic Microwave Background (CMB) - photons once were in
thermal equilibrium with charges which further decoupled and
formed thermal background with blackbody radiation spectrum with temperature $T = 2.7
K$ \cite{penzias}. The information about the density fluctuations $\delta \rho$ at the
decoupling epoch was imprinted in the temperature fluctuations
according to the formula
\bea
\label{delT}
\frac{\delta T}{T} &\propto& \frac{\delta \rho}{\rho} \propto
10^{-5}~.
\eea
\end{itemize}

However, as well as about the past, it is interesting to ask about the future of the
universe. The questions which naturally arise are as follows:
\begin{itemize}
\item What type of future evolution will we generally face?
\item Will the universe expand forever? Will it expand fast,
faster, slower \ldots?
\item Will we face any dramatic change of our future evolution?
\item Is it likely that we face an unexpected end of our future evolution?
\item Is there a barotropic equation of state $p(t) = w\varrho(t)$, $w=$ const.
($p$- the pressure, $\varrho$- the energy density) valid throughout the whole evolution,
or, perhaps $w=w(t)$ so that the pressure can be expanded in series as
\bea
\label{series}
p &=& p_0 + \frac{dp}{d\varrho}\mid_0 (\varrho - \varrho_0) + \frac{1}{2!}
\frac{d^2p}{d\varrho^2}\mid_0 (\varrho - \varrho_0)^2 + O\left[(\varrho -
\varrho_0 )^3\right] \nonumber \\
&=&
p_0 + \frac{\dot{p}_0}{\dot{\varrho}_0} (\varrho - \varrho_0)
+ \frac{1}{2!}
\frac{\ddot{p}_0\dot{\varrho}_0 - \dot{p}_0\ddot{\varrho}_0}{\dot{\varrho}^3}
(\varrho - \varrho_0)^2 + O\left[(\varrho - \varrho_0 )^3\right]~,
\eea
where index "0" refers to a quantity taken at the current moment
of the evolution $t=t_0$.
\item To what extend we are able to determine $w_0 = p_0/\varrho_0$,
$w_0^{,} = (dp/d\varrho) \mid_0$,
\ldots?
\end{itemize}

It is generally believed that these questions may, in general, be addressed
within a more fundamental framework than Einstein's general relativity theory, i.e.,
within the framework of fundamental theories of all physical interactions such as
superstring, brane and M-theory \cite{polchin,polch}.

\section{Empty future, Big-Crunch and phantom-driven Big-Rip}

We start our discussion from the Einstein's field equations for
the homogeneous and isotropic Friedmann universe in the form
(we have assumed that $8\pi G = c =1$)
\bea
\label{rho}
\varrho(t) &=& 3 \left(\frac{\dot{a}^2}{a^2} + \frac{K}{a^2}
\right)~,\\
\label{p}
p(t) &=& - \left(2 \frac{\ddot{a}}{a} + \frac{\dot{a}^2}{a^2} + \frac{K}{a^2}
\right)~,
\eea
where $a(t)$ is the scale factor, $K =0, \pm 1$ is the curvature index. These
two equations contain three unknown functions $a, p, \varrho$. In
order to solve the system one usually assumes the equation of
state of a barotropic type, i.e.,
\bea
\label{eos}
p(t) &=& w\varrho(t)
\eea
with $w=$ const. which leads to the three solutions - each of them
starts with Big-Bang singularity in which $a \to 0$, $\varrho, p \to \infty$
- but only one of them (of $K=+1$) terminates at the second singularity
(Big-Crunch) where $a \to 0$, $\varrho, p \to \infty$ while the
other two ($K=0, -1$) continue to an asymptotic emptiness $\varrho, p \to 0$
for $a \to \infty$. Besides, at least one singularity (e.g. Big-Bang)
appears provided the strong energy conditions of Hawking and
Penrose \cite{he}
\bea
R_{\mu\nu} V^{\mu} V^{\nu} &\geq& 0, \hspace{0.5cm} V^{\mu} - {\rm
a}\hspace{5pt}{\rm timelike}\hspace{5pt} {\rm  vector}~,
\eea
($R_{\mu\nu}$ - Ricci tensor) is fulfilled. In terms of the energy density
and pressure it is equivalent to
\bea
\label{strong}
\varrho + 3p \geq 0, \hspace{0.5cm} \varrho + p \geq 0~.
\eea
From (\ref{rho}) and (\ref{p}) one has
\bea
\label{accel}
\frac{\ddot{a}}{a} &=& - \frac{4\pi G}{3} (\varrho + 3p)~,
\eea
which together with (\ref{strong}) means that
\bea
\ddot{a} &\geq& 0~,
\eea
so that the universe decelerates its expansion.

However, the observations of type Ia supernovae \cite{supernovaeold}
in 1998 gave the evidence for
\bea
\ddot{a} < 0~,
\eea
which means that the universe currently accelerates its expansion.
Taking into account (\ref{accel}) it means that there exists {\it
negative pressure} matter (dark energy, quintessence)
\bea
\label{viol}
p < - \frac{1}{3} \varrho
\eea
in the universe which drives this
acceleration.  More precise fit to the data shows that at least $70\%$ of matter in
the universe now has negative pressure and it is neither the visible
matter, nor the dark matter, which can sum up to the remaining $30\%$.
Some possible sources of negative pressure had
already been suggested before 1998 and they were: the cosmological
constant $w=-1$ \cite{Einstein}, cosmic strings $w=-1/3$, and
domain walls $w=-2/3$ \cite{striwall,AJIII}.

In fact, this result gives more evidence for an asymptotic
emptiness in the future of the universe described by the de Sitter
model in which $\varrho \to 0$ when $a = \exp{[(\sqrt{\Lambda/3})t]} \to \infty$. From
(\ref{viol}) it is obvious that the strong energy condition is
violated.

Last but not least, the most recent observations of type Ia supernovae show
\cite{supernovaenew,riess2004} that the pressure may not only be negative,
but it may also be {\it very strongly negative}. In other words, there is not an
observational barrier onto the amount of negative pressure and it
is also very likely that
\bea
p < - \varrho, \hspace{0.5cm} {\rm or} \hspace{5pt} w<-1~.
\eea
This is called {\it phantom} \cite{phantom,mpd03} and it is the
dark energy of a very large negative pressure which violates all
the remained energy conditions, i.e., the null
\bea
\label{null}
T_{\mu\nu} k^{\mu} k^{\nu} &\geq& 0, \hspace{0.5cm} k^{\mu} - {\rm
a}\hspace{5pt}{\rm null}\hspace{5pt} {\rm  vector}~, \hspace{5pt}
{\rm i.e.,} \hspace{0.5cm} \varrho + p \geq 0~,
\eea
($T^{\mu\nu}$ - an energy-momentum tensor), the weak
\bea
\label{weak}
T_{\mu\nu} V^{\mu} V^{\nu} &\geq& 0, \hspace{0.5cm} V^{\mu} - {\rm
a}\hspace{5pt}{\rm timelike}\hspace{5pt} {\rm  vector}~,\hspace{5pt}
{\rm i.e.,} \hspace{0.5cm} \varrho + p \geq 0, \hspace{0.5cm} \rho \geq 0~,
\eea
and the dominant energy
\bea
\label{dominant}
T_{\mu\nu} V^{\mu} V^{\nu} &\geq& 0, \hspace{0.5cm} T_{\mu\nu} V^{\mu} -
{\rm not}\hspace{5pt} {\rm  spacelike}~,\hspace{5pt}
{\rm i.e.,} \hspace{0.5cm} \mid p \mid \leq \varrho, \hspace{0.5cm}
\varrho \geq 0~.
\eea
The violation of these energy conditions seems to be the most
difficult problem for the models admitting $w \leq -1$ because of
the emergence of both classical and quantum instabilities
\cite{instabilities} and there are some suggestions it should be avoided \cite{kaloper}.
Despite violation of all the energy conditions phantom allows for a new type of future singularity -
Big-Rip which characterizes by a blow-up of both the energy
density and the pressure ($\varrho, p \to \infty$) together with
the infinite size of the universe $a \to \infty$.

In order to make this divergence clear, let us briefly study the
basic properties of phantom cosmology. From the conservation law we have
\bea
\label{gammag0}
\varrho &\propto& a^{-3(w+1)} \hspace{0.5cm} {\rm for}
\hspace{0.5cm} w > -1
\eea
and
\bea
\label{gammal0}
\varrho &\propto& a^{3\mid w+1 \mid} \hspace{0.5cm} {\rm for}
\hspace{0.5cm} w < -1
\hspace{0.5cm} {\rm (phantom)} .
\eea
From Eq. (\ref{gammal0}) it is clear that the singularities appear at
infinite values of the scale factor for phantom cosmologies.
This is also obvious after studying the simplest solution of
(\ref{rho}) for one general fluid $p=w\varrho$ only, which gives
\cite{mpd03}
\bea
\label{aforstan}
a(t) &=& a_0 \mid t \mid^{\frac{2}{3(w+1)}}~,
\eea
for ordinary ($w>-1$) matter, and
\bea
\label{aforphan}
a(t) &=& a_0 \mid t \mid^{-\frac{2}{3\mid w+1 \mid}}~,
\eea
for phantom $w+1 = - \mid w+1 \mid <0$ and
\bea
\varrho &\propto& t^{-2} .
\eea
In other words, taking $w = -4/3$ (phantom) one has $a(t) \to \infty$
and $\varrho \to \infty$ if $t \to 0$, while $a(t) \to 0$ and $\varrho \to 0$ if
$t \to \infty$. On the other hand, in a standard case $w = -2/3$,
for example, one has $a(t) \to 0$ and $\varrho \to \infty$ if $t \to 0$,
while $a(t) \to \infty$ and $\varrho \to 0$ if $t \to \infty$.

It is worth noticing that both the non-phantom matter ($-\varrho < p$) and the phantom
matter ($p < -\varrho$) may be mimicked by a scalar field $\phi$ with some potential
$V(\phi)$ with the effective energy density and pressure
\bea
\label{rhophi}
\varrho &=& \pm \frac{1}{2} \dot{\phi}^2 + V(\phi)~,\\
\label{pphi}
p &=& \pm \frac{1}{2} \dot{\phi}^2 + V(\phi)~,
\eea
where the plus sign refers to the non-phantom matter and the minus sign refers to the phantom.
From the formulas (\ref{rhophi})-(\ref{pphi}), it follows that
phantom can be interpreted as a scalar field with negative
kinetic energy (a ghost).

Another interesting remark can be extracted from the Eqs. (\ref{rho})-(\ref{p}) and
(\ref{gammag0})-(\ref{gammal0}) if we admit shear anisotropy $\sigma_0^2/a^6$ ($\sigma_0=$ const.)
and consider nonisotropic Bianchi type IX
models. Namely, for $w < -1$, the shear anisotropy cannot dominate over the phantom matter
on the approach to a singularity when $a \to \infty$, i.e., we have
\bea
\varrho a^{3 \mid w+1 \mid} &>& \frac{\sigma_0^2}{a^6} \hspace{0.5cm} {\rm for}
\hspace{0.5cm} a \to \infty
\eea
and this prevents the appearance of chaotic behaviour of the phantom cosmologies of the
Bianchi type IX \cite{chaos,mpd03}.

Bearing in mind the fact that the Big-Bang/Big-Crunch singularity
appears for $a \to 0$ while the Big-Rip singularity for $a \to \infty$
one may suspect a kind of duality between the standard matter
($p>0$)/quintessence ($-\varrho<p<0$) models and phantom ($p<-\varrho$)
models which is present in the low-energy-effective superstring theory
\cite{meissner}. Indeed, there is such a duality, called {\it phantom
duality}, which explicitly reads as \cite{mpd03}
\bea
w+1 &\leftrightarrow& -(w+1), \hspace{1.cm}
a \leftrightarrow \frac{1}{a}.
\eea
This duality can easily be seen
if we rewrite the system of equations (\ref{rho})-(\ref{p}) in the form
of the nonlinear oscillator
\bea
\label{oscillator}
\ddot{X} - \frac{D^2}{3} \Lambda X + D(D -1) k X^{1- 2/D} &=&
0 ,
\eea
after introducing the variables
\bea
X &=& a^{D(w)}, \hspace{1.cm} D(w) = \frac{3}{2} (1 + w) .
\eea
It is obvious to notice that Eq. (\ref{oscillator}) preserves its form under the
change
\bea
D &\leftrightarrow& - D~.
\eea
Alternatively, this invariance takes form $(H \equiv \dot{a}/a)$ \cite{lazkoz03}
\bea
H &\leftrightarrow& -H~, \hspace{1.cm} \varrho + p \leftrightarrow - (\varrho + p).
\eea
In fact, there is a richer symmetry of the field equations which
includes brane models called {\it phantom triality} \cite{triality}.

\begin{figure}[h]
\label{BR-BR}
\caption{The typical phantom solution which begins at an initial Big-Rip singularity
($a\to \infty$, $\varrho, p \to \infty$) and terminates at final Big-Rip singularity
($a\to \infty$, $\varrho, p \to \infty$). Similarly as in the case of
Big-Bang-to-Big-Crunch evolution many cycles are presumably admitted.}
\begin{center}
\includegraphics[width=7cm]{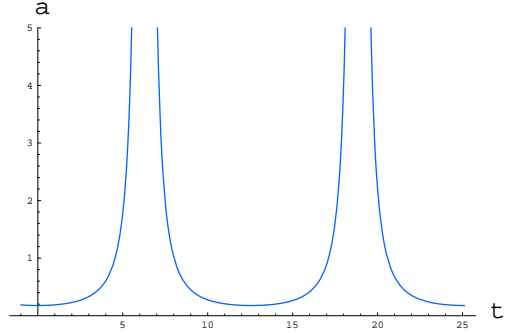}
\end{center}
\end{figure}

The simplest way to consider these dualities is to look at the solutions
(\ref{gammag0})-(\ref{gammal0}). In both cases there is a curvature singularity
at $t=0$, but in the former case it is of a Big-Bang type, while in the latter case it is of
a Big-Rip type. From the observational point of view it is reasonable
to choose the solution (\ref{aforstan}) for positive times $t>0$,
and the solution (\ref{aforphan}) for negative times $t<0$.
Another example of a phantom model with an explicit phantom
duality is (for flat models with walls $w=-2/3$, phantom $w=-4/3$, and $\Lambda < 0$ \cite{mpd03})
\bea
a_w & = & \left(\sin{\frac{\mid D_w \mid^{\frac{1}{2}}}{\sqrt{3}} \mid \Lambda \mid t} \right)^{\frac{1}{D_w}} ,\\
a_{ph} & = & \left(\sin{\frac{\mid D_{ph} \mid^{\frac{1}{2}}}{\sqrt{3}} \mid \Lambda \mid t} \right)^{\frac{1}{D_{ph}}} ,
\eea
where
\bea
D_w = 1/2 = - D_{ph}~,
\eea
so that we have
\bea
a_w &=& a_{ph}^{-1}~.
\eea
It is obvious that the evolution of $a_w$ begins with Big-Bang and
terminates at Big-Crunch while the evolution of $a_{ph}$ begins
with Big-Rip and terminates at Big-Rip (cf. Fig.\ref{BR-BR}).

However, from the point of view of the observations (which support
Hot Big-Bang models), the most interesting models are "hybrid"
models which begin with Big-Bang singularity and terminate at
Big-Rip. Both types of matter (standard and phantom) are present during the
whole evolution of the universe but an early evolution is
dominated by the standard matter, while phantom dominates the late
evolution of it. Of course, this means that there must have been a
change during the evolution from deceleration to acceleration and
this also might have happened quite recently (we will come to this
point later). An explicit example of such an evolution for the
dust ($w=0$) and phantom ($w=-4/3$) model is given in terms of
Weierstrass elliptic functions as
\bea
\left(\frac{a}{a_0}\right)^2 &=& \frac{4 \Omega_{m0}
{\cal P}(\eta)}{4 \Omega_{ph0} {\cal P}^2(\eta) -
\Omega_{m0}}~,\hspace{1.cm}
d\eta = \sqrt{\frac{\Omega_{ph0}a_0}{a}} H_0 dt~,
\eea
where ${\cal P}(\eta)$ is the Weierstrass elliptic function,  $\Omega_{m0}, \Omega_{ph0}$
the density parameters of dust and phantom respectively.

\section{Sudden future singularities of pressure and generalized sudden
future singularities}

Big-Rip may appear in some future time $t=t_{BR}$ analogously to
Big-Crunch (which may appear in some future time $t=t_{BC}$),
but because of growing acceleration it is sometimes
called "sudden". However, we may have something more exotic in the
future evolution of the universe - a singularity which presumably appears
quite unexpectedly and does not violate all the energy conditions.
The hint which allows for such a singularity is that we release
the assumption about the imposition of the equation of state,
i.e., we do not constrain pressure and the energy density in
(\ref{rho}) and (\ref{p}) by any equation like the one in (\ref{eos}).
This enables quite independent time evolution of these physical quantities.

Suppose that we first choose the form of the scale factor as \cite{barrow04}
\bea
\label{sf2}
a(t) &=& A + \left(a_s - A \right) \left(\frac{t}{t_s}\right)^m -
A \left( 1 - \frac{t}{t_s} \right)^n~,
\eea
($A=$ const., $a_s \equiv a(t_s)$) with its time derivatives
\bea
\label{dota}
\dot{a} &=& \frac{m}{t_s} \left(a_s - A\right)
\left(\frac{t}{t_s}\right)^{m-1} + A \frac{n}{t_s}\left( 1 - \frac{t}{t_s}
\right)^{n-1}~,\\
\label{ddota}
\ddot{a} &=& \frac{m\left(m-1\right)}{t_s^2}\left(a_s -A \right)
\left(\frac{t}{t_s}\right)^{m-2}
- A \frac{n(n-1)}{t_s^2}\left( 1 - \frac{t}{t_s} \right)^{n-2}~.
\eea

Choosing
\bea
\label{nq}
1 < n < 2, &\hspace{0.3cm}& 0<m \leq 1~.
\eea
we notice that the scale factor (\ref{sf2})
vanishes and its derivatives (\ref{dota})-(\ref{ddota}) diverge
at $t=0$ leading to a divergence of $\varrho$ and $p$ in
(\ref{rho})-(\ref{p}) (Big-Bang singularity). On the other hand, the scale factor (\ref{sf2})
and its first derivative (\ref{dota}) remain constant while its
second derivative (\ref{ddota}) diverge leading to a divergence of
pressure in (\ref{p}) only, with {\it finite energy density}
(\ref{rho}), i.e.,
\bea
a &=& {\rm const.},\hspace{5pt} \dot{a} = {\rm const.},\hspace{5pt} \ddot{a} \to -\infty,\\
\varrho &=& {\rm const.},\hspace{5pt} p \to \infty,\hspace{5pt} {\rm for}\hspace{5pt} t \to t_s.
\eea
We then conclude that the {\it future singularity} appears and
since it does not show up in the evolution of the scale factor, it is
called "sudden" future singularity (SFS) \cite{barrow04}. This
singularity violates the dominant energy condition
(\ref{dominant}) \cite{lake}, but does not violate any other energy condition.

Sudden future singularities may also be present in the
inhomogeneous models of the universe where they may violate
all the energy conditions. An explicit example of such a
situation was given \cite{sfs1} for the inhomogeneous Stephani universe
\cite{stephani} with no spacetime symmetries with metric
\bea
\label{STMETGEN}
ds^2~&=&~-~\frac{a^2}{\dot{a}^2} \frac{a^2}{V^2}
\left[\left( \frac{V}{a} \right)^{\cdot} \right]^2 dt^2~
+~ \frac{a^2}{V^2} \left[dx^2~+dy^2~+~dz^2 \right]~,
\eea
where
\bea
\label{Vtxyz}
 V(t,x,y,z) =
 1 + \frac{1}{4} k(t) \left\{ \left[x -
x_0(t) \right]^2 + \left[y - y_0(t) \right]^2 + \left[z - z_0(t)
\right]^2 \right\}~,
\eea
and $a, x_0, y_0, z_0$ are arbitrary functions of time. It emerges
that sudden future singularities are similar to finite-density
singularities of pressure (FD) which appear in these Stephani models
since they also do not admit a priori any equation of state.
Besides, sudden future singularities are {\it temporal} pressure
singularities which means that they may appear at an instant $t=t_s$
of the evolution, while finite-density singularities are {\it
spatial} pressure singularities and they may exist somewhere in
the universe nowadays. Sudden future singularities were also proven to
exist in anisotropic models \cite{barrow042}.

If we consider a general time derivative of an order $r$ for
(\ref{sf2}), i.e.,
\bea
\label{dotageneral}
a^{(r)} &=& \frac{m(m-1)(m-2)...(m-r+1)}{t_s^r} \left(a_s -A \right) \left(\frac{t}{t_s}
\right)^{m-r} \nonumber \\
&+& (-1)^{r-1} A \frac{n(n-1)(n-2)...(n-r+1)}{t_s^r}\left( 1 - \frac{t}{t_s} \right)^{n-r}~
\eea
and replace the condition (\ref{nq}) by
\bea
\label{Nnq}
N < n < N+1, &\hspace{0.3cm}& 0<m \leq 1~,
\eea
we realize that for any integer $N+1 = r$ we have a
singularity in the scale factor derivative $a^{(r)}$ (cf.
(\ref{dotageneral})), and consequently in the appropriate pressure derivative $p^{(r-2)}$.
This, for any $r \geq 3$, gives a sudden future singularity which
{\it fulfills all the energy conditions} including the dominant
one \cite{barrow041,lake}. These singularities are, for example, possible in the theories
with higher-order curvature quantum corrections \cite{nojiri}.

In fact, sudden future singularities are determined by a blow-up
of the Riemann tensor and its derivatives rather than the scale
factor (or its inverse) itself. This has an important consequence
on the nature of these singularities in comparison with Big-Bang or
Big-Rip singularities. Namely, the geodesics do not feel the
sudden future singularities at all, since geodesic
equations
\bea
\left(\frac{dt}{d\tau}\right)^2 &=& A + \frac{P^2 +
KL^2}{a(t)}~,\\
\frac{dr}{d\tau} &=& \frac{P_1 cos{\phi} + P_2
\sin{\phi}}{a(t)f(r)}~,\\
\frac{d\phi}{d\tau} &=& \frac{L}{a(t)r^2}~,
\eea
are not singular for $a_s = a(t_s)=$ const. \cite{lazkoz04}.
Here $(t,r,\theta=\pi/2,\phi)$ are coordinates of the Friedmann universe, $\tau$ is the
proper time, $K=0,\pm 1$, $A, L, P^2 = P_1^2 + P_2^2$ are constants, and $f^2(r) =
1/(1-Kr^2)$. On the other hand, the geodesic deviation equation
\bea
\frac{D^2n^{\alpha}}{d\lambda^2} + R^{\alpha}_{~\beta \gamma
\delta} u^{\beta} n^{\gamma} u^{\delta} = 0~,
\eea
where $n^{\alpha}$ is the deviation vector, and $u^{\beta}$ is the
four-velocity vector ($\lambda$ - an affine parameter), feels this
singularity since at $t=t_s$ we have the Riemann tensor $R^{\alpha}_{~\beta \gamma
\delta} \to \infty$. This means point particles do not see sudden
future singularities, but extended objects do in the sense that
they may suffer infinite tidal forces although they may not be torn on crossing
these singularities.

Besides, the spacetimes with sudden future singularities do not
lead to geodesic incompletness - they are {\it weak singularities}
according to the definitions of Tipler \cite{tipler} and Kr\'olak
\cite{krolak}. In other words, sudden future singularities are not
final state of the universe \cite{lazkoz04}. This situation is
somewhat similar to what happens with finite density singularities
in inhomogeneous Stephani models \cite{dabrowski93} although there
instead of the extension of geodesics one has the extension of the
hypersurfaces of constant time throughout these singularities.

In order to conclude, one has to stress that sudden future
singularities and finite density singularities are totally
different from Big-Rip singularities due to phantom since Big-Rip
singularities are felt by geodesic equations and the evolution
cannot be extended behind them. In other words, Big-Rip (like
Big-Crunch) is {\it really} the final state of the evolution while
sudden future singularity is not.

\section{Statefinders and the diagnosis of the future state of the universe}
\label{statef}

From the discussion of the previous sections it is clear that all
the types of singularities (Big-Bang, Big-Crunch, Big-Rip, sudden
future singularities and generalized sudden future singularities)
are related to a blow-up of the scale factor and its time
derivatives
\bea
a(t), \dot{a}(t), \ddot{a}(t), \dddot{a}(t) \ldots~,
\eea
and so to a blow-up of the energy density and pressure derivatives
\bea
\varrho(t), \dot{\varrho}(t), \ldots, p(t), \dot{p}(t), \ldots~.
\eea
Formally, we may translate all that into the observational
characteristics of the expansion (statefinders) such as: the well-known Hubble parameter
\bea
\label{hubb}
H &=& \frac{\dot{a}}{a}~,
\eea
and the deceleration parameter
\bea
\label{dec}
q  &=&  - \frac{1}{H^2} \frac{\ddot{a}}{a} = - \frac{\ddot{a}a}{\dot{a}^2}~,
\eea
together with the new characteristics: the jerk parameter \cite{jerk}
\bea
\label{jerk}
j &=& \frac{1}{H^3} \frac{\dddot{a}}{a} =
\frac{\dddot{a}a^2}{\dot{a}^3}~,
\eea
and the "kerk" (snap) parameter \cite{snap,genphan}
\bea
\label{kerk}
k &=& -\frac{1}{H^4} \frac{\ddddot{a}}{a} =
-\frac{\ddddot{a}a^3}{\dot{a}^4}~,
\eea
the "lerk" parameter
\bea
\label{lerk}
l &=& \frac{1}{H^5} \frac{a^{(5)}}{a} =
\frac{a^{(5)} a^4}{\dot{a}^5}~,
\eea
and "merk", "nerk", "oerk", "perk" etc. parameters, of which a general term
may be expressed as
\bea
\label{dergen}
x^{(i)} &=& (-1)^{i+1}\frac{1}{H^{i}} \frac{a^{(i)}}{a} = (-1)^{i+1}
\frac{a^{(i)} a^{i-1}}{\dot{a}^{i+1}}~,
\eea
and its time derivative reads as
\bea
\label{dergentd}
\left(x^{(i)}\right)^{\cdot} &=& H \left[i(q+1)x^{(i)} - \left( x^{(i+1)} +
x^{(i)} \right) \right]~.
\eea

Future predicted ("sudden") {\it blow-up of statefinders} may easily be linked to an
emergence of future singularities. In particular, this may be the
proper signals for sudden future and generalized sudden future singularities.

The blow-up of statefinders can be read-off redshift-magnitude
relation up to an appropriate order in redshift $z$ after using
the redshift-magnitude relation applied to supernovae data
\cite{snap,genphan}
\bea
\label{m(z)3}
&& m-M = 5\log_{10}{(cz)} - 5\log_{10}{H_0}
\left(\frac{5}{2} \log_{10}{e} \right) \nonumber \\
&& \left\{ (1-q_0) z + \frac{1}{3} \left[\frac{q_0}{2}
\left(\frac{9}{2} q_0 + 5 \right) - j_0 - \frac{7}{4} -
\Omega_{K0} \right] z^2 \right. \\
&& \left. \frac{1}{24} \left[ 2j_0 \left(8q_0 + 5 \right)
- 2k_0 - q_0 \left(7q_0^2 + 11 q_0 + 23 \right) + 25 \right. \right.
\nonumber \\
&& \left. \left. + 4 \Omega_{K0} \left(2 q_0 + 1 \right)
\right]z^3 + O(z^4) \right\}~.\nonumber
\eea
where $m$ is the observed magnitude, $M$ is the absolute magnitude, $c$ is the
velocity of light, and $\Omega_{K0} = K/(H_0^2a_0^2)$.
From (\ref{m(z)3}) it is clear that the jerk appears in the second
order of the expansion and the "kerk" appears in the third order
of this expansion. In particular, the simple signal for
Big-Bang, Big-Crunch or Big-Rip would be $\mid H_0 \mid \to
\infty$; for SFS would be $\mid q_0\mid \to \infty$, while for
generalized SFS would be $\mid j_0\mid \to \infty$, $\mid k_0\mid \to \infty$
etc.

As it was mentioned in the Introduction, statefinders may be
useful in determination of the current status of an equation of
state (\ref{series}) for the matter in the universe.

The most recent analysis of supernovae type Ia data \cite{riess2004} shows that
\bea
j_0 &>& 0~,
\eea
giving no hint about $k_0$. Apart from that it shows that
acceleration of the universe started quite recently at redshift $z=0.46 \pm
0.13$.

\section{SFS avoidance - generalized energy conditions}

From the analysis of the previous Section \ref{statef} we learned that
Big-Crunch singularity in future may emerge despite all the energy
conditions {\it are} fulfilled. On the other hand, Big-Rip may emerge
despite the energy conditions {\it are not} fulfilled. Further,
SFS emerges when only the {\it dominant} energy condition is
violated, while generalized SFS do not lead to any violation of the energy
conditions. This means that an applicability of the standard
energy conditions of Hawking and Penrose \cite{he} to cosmological
models with more exotic properties is limited and so not very
useful.

Because of that, one may consider some more general or different
energy conditions which may be helpful in classification of
singularities in contemporary cosmology based on the
fundamental physical theories such as superstring theory,
brane theory, or M-theory \cite{hw,rs,superjim,veneziano,quevedo,turok0}.

First, let us consider the following higher-order {\it dominant}
energy conditions \cite{statef1}:
\bea
\dot{\varrho} &\geq& 0, \hspace{1.cm} - \dot{p} \leq \dot{\varrho}
\leq \dot{p}~,\\
\ddot{\varrho} &\geq& 0, \hspace{1.cm} - \ddot{p} \leq \ddot{\varrho}
\leq \ddot{p}~,\\
\dddot{\varrho} &\geq& 0, \hspace{1.cm} - \dddot{p} \leq \dddot{\varrho}
\leq \dddot{p}~,\\
\ldots &\ldots& .., \hspace{1.cm} \ldots\ldots\ldots~.\nonumber
\eea
The application of the formulas (\ref{rho}) and (\ref{p}) together
with the following equalities for the time derivatives of the
Hubble parameter (\ref{hubb})
\bea
\label{dotH}
\dot{H} &=& -H^2 \left( q+1 \right)~,\\
\label{ddotH}
\ddot{H} &=& H^3 \left( j+3q+2 \right)~,\\
\label{dddotH}
\dddot{H} &=& -H^4 \left[ k+4j+3q \left( q+4 \right)+6 \right]~,\\
\label{ddddotH}
\ddddot{H} &=& H^5 \left[l+5k+10j(q+2)+30q(q+2)+24 \right]~,
\eea
allows us to write the higher-order energy conditions in terms of
statefinders. The first-order dominant energy condition now reads as
\bea
\label{dom1st+}
\dot{\varrho} + \dot{p} &=& - 2H^3 \left({\rm {\bf j}} + 3q + 2 + 2\frac{K}{a^2H^2}
\right)\geq 0~~,\\
\label{dom1st-}
\dot{\varrho} - \dot{p} &=& - 2H^3 \left(-{\rm {\bf j}} + 3q + 4 + 4\frac{K}{a^2H^2}
\right)\geq 0~~,
\eea
the second-order reads as
\bea
\label{dom2nd+}
\ddot{\varrho} + \ddot{p} &=& 2H^4 \left[{\rm{\bf k}} + 4j + 3q(q+4) + 6
+ 2\frac{K}{a^2H^2}\left(q + 3 \right) \right] \geq 0~~,\\
\label{dom2nd-}
\ddot{\varrho} - \ddot{p} &=& 2H^4 \left[-{\rm {\bf k}} + 2j + 3q(q+6) + 12
+ 4 \frac{K}{a^2H^2}(q+3)\right] \geq 0~~,
\eea
while the third-order reads as
\bea
\label{dom3rd+}
\dddot{\varrho} + \dddot{p} &=& \\
&-& 2H^5 \left[{\rm {\bf l}} + 5k + 10 j (q+2) + 30 q(q+2) + 24
 + 2 \frac{K}{a^2H^2} \left(j + 9q + 12 \right) \right] \geq 0~~,\nonumber \\
\label{dom3rd-}
\dddot{\varrho} - \dddot{p} &=&  \\
&-& 2H^5 \left[-{\rm {\bf l}} + k + 2j (4q + 11) + 6(7q^2 + 17q + 8)
+ 4 \left( j + 9q + 12 \right)\frac{K}{a^2H^2}
\right] \geq 0~, \nonumber
\eea
and so on. From these relations one can easily see that it is {\it not
possible to fulfill} any generalized dominant energy condition if
any of statefinders $j$, $k$, $l$, etc. is singular. This is due
to the fact that the signs in the appropriate expressions in front
of statefinders are the opposite. This gives a conclusion that the
violation of the higher-order dominant energy conditions can be
good signals for the emergence of the generalized sudden future
singularities.

In a similar manner one is able to consider the higher-order null, weak
and strong energy conditions as follows
\bea
\label{null1st}
\dot{\varrho} + \dot{p} &=& - 2H^3 \left({\rm {\bf j}} + 3q + 2 + 2\frac{K}{a^2H^2} \right)\geq
0~,\\
\label{weak1st}
\dot{\varrho} + \dot{p} &=& - 2H^3 \left({\rm {\bf j}} + 3q + 2 + 2\frac{K}{a^2H^2} \right)\geq 0~,
\hspace{0.5cm} \dot{\varrho} \geq 0~,\\
\label{strong1st}
\dot{\varrho} + 3 \dot{p} &=& - 6 H^3 ({\rm {\bf j}} + q)\geq 0~,\hspace{0.5cm}
\dot{\varrho} + \dot{p} = - 2H^3 \left({\rm {\bf j}} + 3q + 2 +
2\frac{K}{a^2H^2} \right)\geq 0~, \\
\ldots + \ldots &=& \ldots \ldots \geq 0 \nonumber ~,
\eea
and so on.

After having a closer look at them it emerges that these higher-order energy
conditions are not always very useful in order to determine the future fate of the Universe.

However, some hybrid energy conditions may work for that.
For example an energy condition \cite{barrow042}
\bea
\alpha \varrho &>& \dot{p}~,
\eea
with $\alpha=$ const., which in terms of statefinders gives
\bea
j &>& \left(1 - \frac{3\alpha}{2H} \right)\left(1 + \frac{K}{a^2H^2}
\right)~,
\eea
may prevent the emergence of sudden future singularity
singularity for $N=2$ in (\ref{Nnq}).

\section{Conclusions}

In view of the discussion performed in this paper one has the
following remarks.

Firstly, the future state of the universe may be {\it more sudden
and violent}. It means that the universe may terminate either in a sudden
future singularity or in a Big-Rip singularity and this is totally
different from our earlier expectations that it could terminate in an
asymptotically de Sitter state or in a Big-Crunch.

Secondly, these new future singularities (Big-Rip and SFS) should
not be confused - they have totally different properties with
respect to geodesic completness. In particular, one can extend the
evolution of the universe through a sudden future singularity, while
it is not possible to do so for a Big-Rip.

Thirdly, statefinders (Hubble, deceleration, jerk, kerk etc.) may
be useful to diagnose the future state of the universe. By this we
mean the emergence of sudden future singularities, the emergence
of generalized sudden future singularities, the evolution of the
cosmic equation of state $w = w(t)$ etc.

Finally, the new energy conditions may be introduced for the sake
of the proper signal for generalized sudden future singularities, or
(on the contrary), for the sake of the avoidance of sudden future singularities, or
generalized sudden future singularities.

\begin{acknowledgement}
This work was partially supported by the Polish Ministry of
Education and Science grant No 1 P03B 043 29 (years 2005-2007).
\end{acknowledgement}




\end{document}